\documentclass[twocolumn,showpacs,preprintnumbers,amsmath,amssymb]{revtex4}
\usepackage{graphicx}% Include figure files
\usepackage{dcolumn}% Align table columns on decimal point
\usepackage{bm}% bold math

\begin{document}

\preprint{Mag-nano-2002}

\title{Magnetic nanographite}% Force line breaks with \\

\author{Koichi Kusakabe}
 \email{kabe@sunshine.gs.niigata-u.ac.jp}
\author{Masanori Maruyama}
 \email{maruyama@sunshine.gs.niigata-u.ac.jp}
\affiliation{%
Graduate School of Science and Technology, 
Niigata University, \\
Ikarashi, Niigata 950-2181, Japan
}%

\date{\today}% It is always \today, today,

\begin{abstract}
Hydrogenated nanographite can 
display spontaneous magnetism. 
Recently we proposed that 
hydrogenation of nanographite is able to induce 
finite magnetization. 
We have performed theoretical investigation of 
a graphene ribbon in which 
each carbon is bonded to two hydrogen atoms at one edge 
and to a single hydrogen atom at another edge. 
Application of the local-spin-density approximation 
to the calculation of the electronic 
band-structure of the ribbon shows 
appearance of a spin-polarized flat band at the Fermi energy. 
Producing different numbers of mono-hydrogenated carbons and 
di-hydrogenated carbons can create magnetic moments in nanographite. 
\end{abstract}

\maketitle

%\section{\label{sec:level1}First-level heading:\protect\\ The line
%break was forced \lowercase{via} \textbackslash\textbackslash}

%Introduction part

Recent reports on magnetic carbon have stimulated renewed interest 
in carbon systems as possible new magnetic materials 
made exclusively of light elements. 
These investigations include experiments suggesting ferromagnetic behavior 
in polymerized fullerenes made 
using photo-assisted oxidation\cite{Murakami96} 
or high-pressure treatment.\cite{Makarova01} 
A nanometer-scale graphite called nanographite may yield 
another magnetic carbon system. 
Some experiments have demonstrated the 
existence of anomalous magnetic behavior unexpected in 
diamagnetic bulk graphite.\cite{Shibayama00,Prasad00,Kopelevich00} 

Several theoretical investigations of nanographite 
have been carried out prior to or parallel to these 
experiments.\cite{Fujita96,Nakada96,Fujita97,Wakabayashi98,Miyamoto99,Harigaya01,Harigaya01-2,Okada01,Harigaya02} 
An important finding was that $\pi$ electrons in 
nanometer-scale graphite are strongly affected 
by structure of edges. 
%%%Revised
A typical edge is the zigzag edge. 
The zigzag edge was found in 
STM images of nanometer scale graphene,\cite{Land,Nagashima} 
although the armchair edge has lower 
energy than the zigzag edge.\cite{Thess,Lee} 
Fujita and coworkers 
suggested that the $\pi$ electrons on a mono-hydrogenated zigzag edge 
may create a ferrimagnetic spin structure on the edge.\cite{Fujita96} 
%%%
Magnetism with localized spin moments is possible due to 
the existence of non-bonding localized states at the zigzag edge. 
These states are called edge states and are highly degenerate 
at the Fermi level and may be spin polarized. 
Such an edge state does not appear on an armchair edge. 
Klein has studied another edge of graphene strips, 
considering a zigzag edge with an atomic site 
having a $\pi$-orbital on the edge.\cite{Klein,Klein-Bytautas} 
Examining a tight-binding model of $\pi$ electrons, 
he found degenerate surface states on the edge, 
which is referred to here as a bearded edge. 

The total spin moment of a graphene ribbon, however, should be zero, 
if both of the two edges of the ribbon are zigzag edges, 
or if both of them are bearded edges. 
This is because, the structure of the $\pi$ network 
becomes a bipartite lattice 
having the same number of sublattice sites. 
In such cases, an argument using the Hubbard model 
results in zero spin moment, 
following the Lieb theorem.\cite{Fujita96,Lieb} 
Naively, we may say that 
local magnetic moments at the two edges are coupled 
anti-ferromagnetically.\cite{Fujita96} 

%Motivation and purpose

In this paper, we show that a hydrogenated graphene ribbon 
can have a finite total magnetic moment. 
A key point is the structure of the two edges of the ribbon. 
In our ribbon, one of the edges is composed of 
mono-hydrogenated carbon atoms and another is 
made of di-hydrogenated ones. 
By a first-principles electronic-state calculation, 
we will show that the system is stable and has 
a fully spin-polarized flat band. A finite spontaneous 
magnetization appears in the graphene ribbon. 

We present here our idea to find magnetic nanographite. 
If a graphene ribbon has 
a bearded edge on one side and a zigzag edge on the other side, 
the $\pi$ network becomes a bipartite lattice with different numbers of 
sublattice sites.\cite{Wakabayashi,Kusakabe-Takagi} 
Then, a flat band appears at the Fermi level. 
The band is fully spin polarized if inter-electron 
interactions {\it e.g.} the Hubbard repulsion $U$ is considered.\cite{Lieb} 
In addition, the flat band consists of edge states, so that 
the magnetic moments are localized at both edges 
and the distribution of moments shows exponential decay 
in the $\pi$ network.\cite{Kusakabe-Takagi} 
However, to design a realistic material, 
we also have to specify the atomic configuration of the bearded edge. 

We found that 
one solution is a zigzag edge whose end carbon atoms are 
di-hydrogenated.\cite{Kusakabe-Maruyama} 
In order to present an example of magnetic nanographite, 
we will show 
the electronic band structure of the ribbon which has 
a completely spin-polarized flat band at the Fermi energy. 
The magnetism will be explained using a Hubbard model.

%Methods of calculation

To investigate the stability and magnetic structure 
of the nanographite, we used an electronic-state calculation 
with a local spin-density approximation (LSDA) based 
on the density functional theory.\cite{Hohenberg-Kohn,Kohn-Sham} 
An LSDA functional given by Perdew and Wang\cite{LDAPW92} was utilized 
as the exchange-correlation energy functional. 
The ultra-soft pseudopotential generated using the Vanderbilt strategy 
was adopted.\cite{Vanderbilt} 
The valence wave functions 
were expanded in a plane-wave (PW) basis set. 
To achieve convergence in the total energy, 
we used an energy cut-off $E_c$ of 49Ry 
for structural optimization in a fixed unit cell. 
To reduce computation time, 
we adopted $E_c=25$Ry to optimize the size of the unit cell 
before the final optimization of inner coordinates. 
Structural optimization was performed until each component of the 
interatomic force became less than $8\times 10^{-5}$ H/a.u. 
The calculations were done using a computer program 
called the Tokyo ab-initio program package (TAPP).\cite{Yamauchi} 

% Check of the pseudopotential

To check the accuracy of our calculation, we obtained 
the structural parameters of a hydrogen molecule, a CH$_4$ molecule, 
a C$_2$H$_4$ molecule, diamond and graphite. 
The bond length of each molecule 
and the bond angles of CH$_4$ and C$_2$H$_4$ were obtained 
with relative errors less than 3.3\%. 
The vibrational energy $\hbar\omega_e$ of H$_2$ was $1.92 \times 10^{-2}$ [H], 
which is about 4\% smaller than $2.00\times 10^{-2}$ [H] observed. 
For cubic diamond, the lattice constant, the bulk modulus and its 
pressure derivative were 6.677 a.u., 4.66 Mbar and 3.68. 
For hexagonal graphite, the lattice constants were 
$a=4.616$ a.u. and $c=12.67$ a.u. 
These values are in reasonable agreement with the values 
determined experimentally or by previous calculations.\cite{Furthmuller} 
Band structures of the solids also reproduced previous 
results.\cite{Furthmuller} 

% Structure with hydrogen terminated edge

\begin{figure}
\includegraphics[height=2.3in]{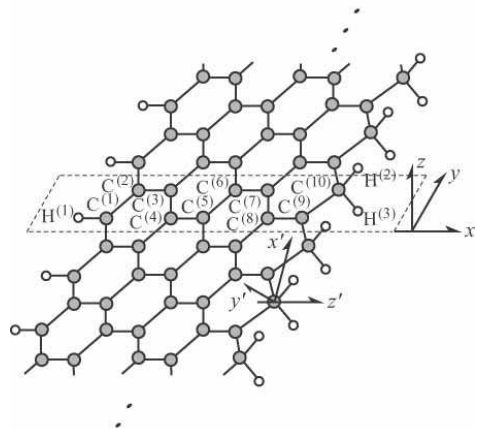}% Here is how to import EPS art
\caption{\label{Structuremodel} 
A structural model of a magnetic graphene ribbon. 
Dark circles represent carbon atoms and open circles denote 
hydrogen atoms. Dashed lines represent a section of a unit cell 
in the $xy$-plane. The unit cell contains $2n$ carbon atoms 
and 3 hydrogen atoms. Here a case with $n=5$ is shown. 
Two sets of 3-dimensional coordinate axes, $(xyz)$ and $(x'y'z')$, 
are displayed. 
%%%Revised
The $x$-axis and the $y$-axis are in a graphene plane. 
Three axes $x'$, $y'$ and $z'$ point to the (011), (0-11), 
(100) directions, respectively, in the $(xyz)$ coordinate. 
%%%
}
\end{figure}

Consider the graphene ribbon shown in Fig. \ref{Structuremodel}. 
In this structure, each dark circle represents 
the position of a carbon atom. Solid lines represent 
$\sigma$ bonds. 
Each open circle is a hydrogen atom. 
For convenience, we number atoms 
and name them C$^{(1)}$ $\cdots$ C$^{(2n)}$ 
and H$^{(1)}$ $\cdots$ H$^{(3)}$ in a unit cell. 
Here, $n$ is a positive integer. 
On the left edge, each carbon atom 
is bonded to a hydrogen atom and has a $\pi$ orbital. 
On the right edge, each carbon atom has bonds to 
two hydrogen atoms and two neighboring carbon atoms. 
If we consider that 
the di-hidrogenated carbon atom C$^{(2n)}$ forms 
sp$^3$ bonds instead of sp$^2$ bonds, 
C$^{(2n)}$ loses a $\pi$ electron. 
Since there remain $2n-1$ carbon atoms having a $\pi$ electron 
per each in a unit cell, the $\pi$ network might become 
a bipartite lattice with different numbers of sublattice sites. 

We can consider ribbons with various widths. 
However, we know that the optimum width is given by $n\simeq 5 \sim 10$ 
for nanographite ribbons with zigzag edges.\cite{Nakada96} 
In any case, the essential properties should be the same 
irrespective of the width. 
%%%Revised
Thus we first consider a ribbon with $n=5$ 
and discuss dependence of magnetism on the width of the 
graphene ribbon at the end of the paper. 
%%%

To examine our idea, we have performed an LSDA calculation. 
Below, we assume a periodic boundary condition. 
%%% Revised
Since the structure is one-dimensional, we are allowed to use 
$k$ points along the $y$ axis as $(0,k_y,0)$ for the structural optimization. 
We used 6 $k$ points uniformly distributed 
in the Brillouin zone for most simulations. 
We determined the equilibrium lattice constant of the ribbon, 
which became 4.674 a.u. 
For the calculation to optimize the inner parameters, 
the size of the cell was taken 
to be 40.27 $\times$ 4.647 $\times$ 18.60 (a.u.)$^3$. 
%%%
The spacing between neighboring ribbons became 
19.5 a.u. in the $x$ direction and 15.3 a.u. in the $z$ direction. 
The structure was optimized by a conjugate gradient method. 
%%% Revised
The obtained results did not change 
when we used 12 $k$ points instead. 
%%%
The bond length of C-C ranged from about 2.64 $\sim$ 2.70 a.u. 
for neighboring carbon pairs except for 2.59 a.u. 
between C$^{(8)}$ and C$^{(9)}$ and 
2.76 a.u. between C$^{(9)}$ and C$^{(10)}$. 
Bond angles H$^{(2)}$-C$^{(10)}$-H$^{(3)}$ 
and C$^{(9)}$-C$^{(10)}$-C$^{(9)}$ 
are 100.7$^\circ$ and 114.8$^\circ$. 
These structure parameters suggest that the end carbon 
C$^{(10)}$ forms sp$^3$ bonds.

%Band structure

\begin{figure}
\includegraphics[width=3.0in]{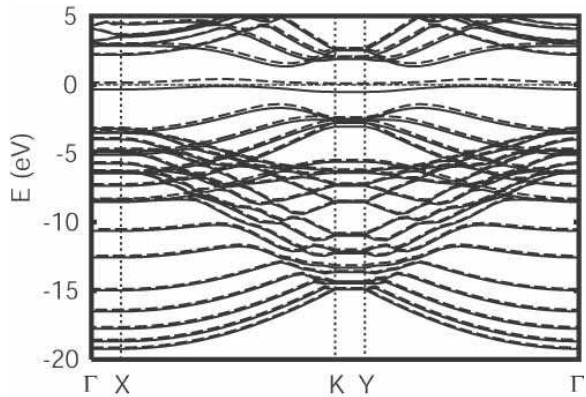}% Here is how to import EPS art
\caption{\label{Bandstructure} 
The electronic band structure of a hydrogenated graphene ribbon. 
Solid lines and dashed lines represent 
the dispersion relation of bands for 
up electrons and down electrons, respectively. 
At the Fermi level $E_F=0$, a dispersionless band appears. 
The band is completely spin polarized. 
Above and blow the flat band, there are 8 $\pi$-bands, 
which are from the 18th band to the 21th band and from 
the 23th band to the 26 th band on the $K$-$Y$ line. 
The 17th band on the $K$-$Y$ line is composed mainly of 
1s orbitals of two hydrogen atoms H$^{(2)}$ and H$^{(3)}$ in Fig. 1. }
\end{figure}

An electronic band structure of the graphene ribbon 
is shown in Fig. \ref{Bandstructure}. 
In the first Brillouin zone, 
the $\Gamma$-Y line corresponds to a one-dimensional 
axis on which a $k$ vector is parallel to the ribbon. 
A flat band appears at the Fermi level. 
The band is almost dispersionless in the whole Brillouin zone and 
is completely spin polarized. 
That is, only the spin-up branch is filled and 
the spin-down branch is empty. 
Splitting between polarized flat bands for up-electrons and down-electrons 
is approximately 0.5 $\sim$ 0.6 eV and the gap between them is $\sim 0.2$ eV. 
These values give a rough energy scale of stiffness for 
the magnetic ground state. 
The total spin $S_{tot}$ becomes 1/2 times number of unit cells $N_c$. 
%%% Revised
Stability of the polarized flat band was confirmed 
by using a unit cell which was doubled in the $y$ direction. 
%%%

\begin{figure}
\includegraphics[width=3.0in]{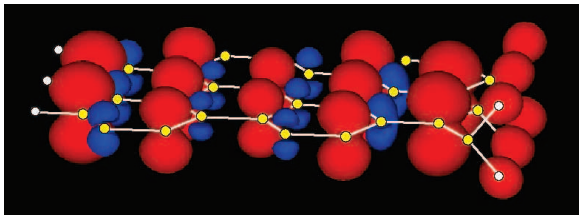}% Here is how to import EPS art
\caption{\label{Spindensity} 
The optimized structure of a magnetic graphene ribbon. 
Yellow spheres and white spheres represent 
the positions of carbon atoms and hydrogen atoms, respectively. 
Spin density on the ribbon is displayed by isosurfaces. 
Red surfaces and blue surfaces represent spin-up density and 
spin-down density. The isolevel is set to 
1/4 of the maximum amplitude of the spin-up density. 
}
\end{figure}

The flat band is the center band of 
9 well-characterized $\pi$-bands. See band dispersion 
around the $Y$ point in Fig. \ref{Bandstructure}. 
The dispersion relation of these $\pi$ bands is roughly the same as 
those obtained for a simple tight binding model 
with 9 $\pi$-orbitals on a graphene ribbon. 
In Fig. \ref{Bandstructure}, all branches for up-electrons 
and down-electrons are 
split by spin polarization. 
This result implies that a magnetic moment 
appears on every atom. 
%%% Revised
We show obtained spin density in Fig. \ref{Spindensity}. 
A ferrimagnetic spin structure is clearly seen. 
A local spin moment exists on every carbon atom. 
Interestingly, finite spin moments appear 
on two hydrogen atoms which have bondings with C$^{(10)}$. 
The reason will be explained below. 
%%%
Thus, the graphene ribbon is confirmed to be magnetic 
in the LSDA model. 
We call this structure magnetic nanographite.

%%% Revised
%%%

%Wave function

We found a curious characteristic of the wave functions of the flat band. 
We surmise that only 9 carbon atoms in the unit cell possess 
$\pi$ orbitals which contribute to the $\pi$ bands. 
This is because sp$^3$ bonds are created 
at the di-hydrogenated carbon atom C$^{(10)}$. 
Wave functions on the flat band were non-bonding edge states 
with finite amplitude on only odd-numbered carbon atoms, but not on 
even-numbered ones. 
An interesting point is that a finite amplitude was found on 
two hydrogen atoms H$^{(2)}$ and H$^{(3)}$. 
Therefore, we have to consider 1s orbitals on these hydrogen atoms 
in order to analyze the spin-polarized flat band. 

%sp^3 -> a pi orbital

Let us introduce two sets of orthogonal coordinates, $(xyz)$ and $(x'y'z')$ 
as defined in Fig. \ref{Structuremodel}. 
Four sp$^3$ orbitals on C$^{(10)}$ may be represented as 
$\phi_1 = \phi_s + \phi_{x'} + \phi_{y'} + \phi_{z'}$, 
$\phi_2 = \phi_s - \phi_{x'} - \phi_{y'} + \phi_{z'}$, 
$\phi_3 = \phi_s + \phi_{x'} - \phi_{y'} - \phi_{z'}$, and 
$\phi_4 = \phi_s - \phi_{x'} + \phi_{y'} - \phi_{z'}$.
Here, $\phi_s$ is a 2s orbital and $\phi_m$ ($m=x',y',z'$) are 
three 2p orbitals on C$^{(10)}$. 
$\phi_1$ and $\phi_2$ point to H$^{(2)}$ and H$^{(3)}$, respectively. 
We take linear combinations of $\phi_1$ and $\phi_2$ as well as 
1s orbitals $\varphi_1$ and $\varphi_2$ on H$^{(2)}$ and H$^{(3)}$ 
and define $\bar{\phi}_\pm=(\phi_1\pm\phi_2)/\sqrt{2}$ and 
$\bar{\varphi}_\pm=(\varphi_1\pm\varphi_2)/\sqrt{2}$. 
Note that $\bar{\phi}_-=\phi_z$, {\it i.e.} a $\pi$ orbital 
on C$^{(10)}$. By symmetry, $\bar{\phi}_-$ and $\bar{\varphi}_-$ 
hybridize with the $\pi$ orbitals of the host $\pi$ system 
of the graphene sheet, but not with $\sigma$ orbitals 
in the honeycomb lattice. 
Therefore, we must utilize 
$\bar{\phi}_-$ and $\bar{\varphi}_-$ as well as 
the $\pi$ orbitals on C$^{(i)}$ ($i=1,\cdots 9$) 
to analyze the $\pi$ network in the magnetic nanographite.

%Analysis using the Hubbard model

If we adopt the above representation, 
we obtain an effective model describing the ribbon. 
The model consists of 11 local orbitals in a unit cell. 
Here we ignore overlap integrals between orbitals for simplicity. 
Introducing the operator representation $c_{i,\sigma}$ 
for a $\pi$ electron on a carbon atom and 
$d_{l,\sigma}$ for an electron in $\bar{\varphi}_-$ 
in the $l$-th unit cell, 
we obtain an effective Hamiltonian. 
\begin{eqnarray}
H&=&-\sum_{\langle i,j \rangle}\sum_\sigma t_{i,j} 
(c^\dagger_{i,\sigma} c_{j,\sigma} + H.c. ) 
+\sum_i U n_{i,\uparrow} n_{i,\downarrow} \nonumber \\
& &-\sum_{\langle i,l \rangle}\sum_\sigma t'_{i,l} 
(c^\dagger_{i,\sigma} d_{l,\sigma} + H.c. ) 
+\sum_l U' m_{l,\uparrow} m_{l,\downarrow} \nonumber \\
& &+ \varepsilon \sum_l \sum_\sigma m_{l,\sigma}
\end{eqnarray}
Here, an index $i$ specifying a carbon atom runs over all carbon sites. 
Two kinds of number operators are defined as 
$n_{i,\sigma}\equiv c^\dagger_{i,\sigma}c_{i,\sigma} $
and $m_{l,\sigma}\equiv d^\dagger_{l,\sigma}d_{l,\sigma} $. 
Bond connections given by non-zero $t_{i,j}$ and $t'_{i,l}$ 
are the same as those in a graphene ribbon 
with a zigzag edge and a bearded edge.\cite{Kusakabe-Takagi} 
We set the origin of the energy at the orbital energy of 
the 2p orbital 
$\varepsilon_{2p}$ on carbon. 
$\varepsilon_{1s}$ denotes the energy 
of a 1s electron on a hydrogen atom and 
$\varepsilon$ is given by 
$\varepsilon = \varepsilon_{1s}-\varepsilon_{2p}$. 
Since $\varepsilon <0$, 
a band mainly composed of $\bar{\varphi}_-$ 
is separated in energy from other $\pi$ bands. 
This property is also seen in the LSDA band structure, where 
the 17th band on the $K$-$Y$ line in Fig. \ref{Bandstructure} 
is the corresponding band. 

Consider a periodic boundary condition for a finite system 
with $N_c$ unit cells. 
If $\varepsilon=0$ and if $U'=U$, 
Lieb's theorem is applicable and 
the ground state becomes a ferrimagnetic state with 
a total magnetization $S_{\rm tot} = |N_A-N_B|/2$. 
Here the difference in sublattice sites $|N_A-N_B|$ equals $N_c$. 
This magnetic ground state is thought to be stable against 
perturbation like $\varepsilon<0$, 
since 1) there is a finite gap 
between the flat band and other $\pi$ bands both in 
the tight-binding description and in the LSDA result
in Fig. \ref{Bandstructure}, 
2) we have confirmed that 
a small difference of $\varepsilon_{1s}$ and 
$\varepsilon_{2p}$ does not break the flatness of 
the center band in the tight-binding model. 
Thus the magnetism found in an LSDA calculation can be understood 
as the appearance of ferrimagnetism in a flat-band Hubbard model.

%Formation energy of >C- >C-H >C<H_2

The formation energy of 
a di-hydrogenated edge has been evaluated 
by comparing the total energy within LSDA. 
We denote $E_{H_2}$, $E_{G1}$ and $E_{G2}$ as 
the energy of a hydrogen molecule, that of 
a graphene ribbon with mono-hydrogenated zigzag edges 
and that of a magnetic nanographite ribbon, respectively. 
The latter two are energy per a unit cell including $2n$ carbon atoms. 
Our calculation shows that $E_{G2}-(E_{G1}+E_{H_2}/2) \simeq 0.3$eV 
for $n=5$. 
Thus, the hydrogenation of nanographite with mono-hydrogenated 
zigzag edges is confirmed to be an endothermic reaction. 
Utilization of proper catalysts in catalytic reduction may 
be important in the synthesis of 
magnetic nanographite. 

%%% Effect of the width
%%% Revised
To see how our conclusion is affected by changing 
the width of the ribbon, 
we performed the LSDA calculation for 
ribbons with a width from $n=2$ to $n=4$. 
A ferrimagnetic structure was found 
even in these narrow ribbons. 
These ribbons do not show the Peierls distortion, 
which is similar to conclusions 
on the mono-hydrogenated zigzag ribbons.\cite{Kertesz,Fujita97,Miyamoto99}
%%%

%Summary and discussion

We have shown that a spin-polarized flat band appears 
in the LSDA band structure of a graphene ribbon. 
The origin of the magnetism is explained as a 
realization of the flat-band ferromagnetism. 
An orbital $\bar{\varphi}_-$ on two hydrogen atoms 
H$^{(2)}$ and H$^{(3)}$ is 
an anti-bonding orbital and has symmetry $p_z$. 
Thus, hydrogenation of a carbon atom 
is identified as the addition of a $\pi$ orbital 
at a zigzag edge. 
The structure may be regarded as a realization of the bearded edge. 
%%%Revised
We can expect magnetism for any hydrogenated ribbon with $|N_A-N_B|>0$, 
even if di-hydrogenation of the zigzag edge is not perfect. 
%%%
The same effect is also expected 
with fluorination.\cite{Maruyama-Kusakabe}
In this report we have considered a ribbon 
with a flat graphene structure, 
but the method is also 
applicable to nanotubes with zigzag edges. 
Thus it appears that, 
a magnetic nanotube may be designed.\cite{Kusakabe-Maruyama}

\begin{acknowledgments}
One of the authors K.K. is grateful for fruitful discussions 
with Professor T. Enoki and Dr Takai. 
This work has been supported by a Grant-in-Aid from the Ministry of 
Education, Culture, Sports, Science and Technology of Japan. 
The calculation was partly performed at the 
computer facility of ISSP, University of Tokyo. 
\end{acknowledgments}

\bibliography{Mag-nano}% Produces the bibliography via BibTeX.

\begin{thebibliography}{32}
\expandafter\ifx\csname natexlab\endcsname\relax\def\natexlab#1{#1}\fi
\expandafter\ifx\csname bibnamefont\endcsname\relax
  \def\bibnamefont#1{#1}\fi
\expandafter\ifx\csname bibfnamefont\endcsname\relax
  \def\bibfnamefont#1{#1}\fi
\expandafter\ifx\csname citenamefont\endcsname\relax
  \def\citenamefont#1{#1}\fi
\expandafter\ifx\csname url\endcsname\relax
  \def\url#1{\texttt{#1}}\fi
\expandafter\ifx\csname urlprefix\endcsname\relax\def\urlprefix{URL }\fi
\providecommand{\bibinfo}[2]{#2}
\providecommand{\eprint}[2][]{\url{#2}}

\bibitem[{\citenamefont{Murakami and Suematsu}(1996)}]{Murakami96}
\bibinfo{author}{\bibfnamefont{Y.}~\bibnamefont{Murakami}} \bibnamefont{and}
  \bibinfo{author}{\bibfnamefont{H.}~\bibnamefont{Suematsu}},
  \bibinfo{journal}{Pure Appl. Chem.} \textbf{\bibinfo{volume}{68}},
  \bibinfo{pages}{1463} (\bibinfo{year}{1996}).

\bibitem[{\citenamefont{Makarova et~al.}(2001)\citenamefont{Makarova,
  Sundqvist, Hohne, Esquinazi, Kopelevich, Scharff, Davydov, Kashevarova, and
  Rakhmanina}}]{Makarova01}
\bibinfo{author}{\bibfnamefont{T.~L.} \bibnamefont{Makarova}},
  \bibinfo{author}{\bibfnamefont{B.}~\bibnamefont{Sundqvist}},
  \bibinfo{author}{\bibfnamefont{R.}~\bibnamefont{Hohne}},
  \bibinfo{author}{\bibfnamefont{P.}~\bibnamefont{Esquinazi}},
  \bibinfo{author}{\bibfnamefont{Y.}~\bibnamefont{Kopelevich}},
  \bibinfo{author}{\bibfnamefont{P.}~\bibnamefont{Scharff}},
  \bibinfo{author}{\bibfnamefont{V.~A.} \bibnamefont{Davydov}},
  \bibinfo{author}{\bibfnamefont{L.~S.} \bibnamefont{Kashevarova}},
  \bibnamefont{and} \bibinfo{author}{\bibfnamefont{A.~V.}
  \bibnamefont{Rakhmanina}}, \bibinfo{journal}{Nature}
  \textbf{\bibinfo{volume}{413}}, \bibinfo{pages}{716} (\bibinfo{year}{2001}).

\bibitem[{\citenamefont{Shibayama et~al.}(2000)\citenamefont{Shibayama, Sato,
  Enoki, and Endo}}]{Shibayama00}
\bibinfo{author}{\bibfnamefont{Y.}~\bibnamefont{Shibayama}},
  \bibinfo{author}{\bibfnamefont{H.}~\bibnamefont{Sato}},
  \bibinfo{author}{\bibfnamefont{T.}~\bibnamefont{Enoki}}, \bibnamefont{and}
  \bibinfo{author}{\bibfnamefont{M.}~\bibnamefont{Endo}},
  \bibinfo{journal}{Phys. Rev. Lett.} \textbf{\bibinfo{volume}{84}},
  \bibinfo{pages}{1744} (\bibinfo{year}{2000}).

\bibitem[{\citenamefont{Prasad et~al.}(2000)\citenamefont{Prasad, Sato, Enoki,
  Hishiyama, Kaburagi, Rao, Eklund, Oshida, and Endo}}]{Prasad00}
\bibinfo{author}{\bibfnamefont{B.~L.~V.} \bibnamefont{Prasad}},
  \bibinfo{author}{\bibfnamefont{H.}~\bibnamefont{Sato}},
  \bibinfo{author}{\bibfnamefont{T.}~\bibnamefont{Enoki}},
  \bibinfo{author}{\bibfnamefont{Y.}~\bibnamefont{Hishiyama}},
  \bibinfo{author}{\bibfnamefont{Y.}~\bibnamefont{Kaburagi}},
  \bibinfo{author}{\bibfnamefont{A.~M.} \bibnamefont{Rao}},
  \bibinfo{author}{\bibfnamefont{P.~C.} \bibnamefont{Eklund}},
  \bibinfo{author}{\bibfnamefont{K.}~\bibnamefont{Oshida}}, \bibnamefont{and}
  \bibinfo{author}{\bibfnamefont{M.}~\bibnamefont{Endo}},
  \bibinfo{journal}{Phys. Rev. B} \textbf{\bibinfo{volume}{62}},
  \bibinfo{pages}{11209} (\bibinfo{year}{2000}).

\bibitem[{\citenamefont{Kopelevich et~al.}(2000)\citenamefont{Kopelevich,
  Esquinazi, Torres, and Moehlecke}}]{Kopelevich00}
\bibinfo{author}{\bibfnamefont{Y.}~\bibnamefont{Kopelevich}},
  \bibinfo{author}{\bibfnamefont{P.}~\bibnamefont{Esquinazi}},
  \bibinfo{author}{\bibfnamefont{J.~H.~S.} \bibnamefont{Torres}},
  \bibnamefont{and}
  \bibinfo{author}{\bibfnamefont{S.}~\bibnamefont{Moehlecke}},
  \bibinfo{journal}{J. Low Temp. Phys.} \textbf{\bibinfo{volume}{119}},
  \bibinfo{pages}{691} (\bibinfo{year}{2000}).

\bibitem[{\citenamefont{Fujita et~al.}(1996)\citenamefont{Fujita, Wakabayashi,
  Nakada, and Kusakabe}}]{Fujita96}
\bibinfo{author}{\bibfnamefont{M.}~\bibnamefont{Fujita}},
  \bibinfo{author}{\bibfnamefont{K.}~\bibnamefont{Wakabayashi}},
  \bibinfo{author}{\bibfnamefont{K.}~\bibnamefont{Nakada}}, \bibnamefont{and}
  \bibinfo{author}{\bibfnamefont{K.}~\bibnamefont{Kusakabe}},
  \bibinfo{journal}{J. Phys. Soc. Jpn.} \textbf{\bibinfo{volume}{65}},
  \bibinfo{pages}{1920} (\bibinfo{year}{1996}).

\bibitem[{\citenamefont{Nakada et~al.}(1996)\citenamefont{Nakada, Fujita,
  Dresselhaus, and Dresselhaus}}]{Nakada96}
\bibinfo{author}{\bibfnamefont{K.}~\bibnamefont{Nakada}},
  \bibinfo{author}{\bibfnamefont{M.}~\bibnamefont{Fujita}},
  \bibinfo{author}{\bibfnamefont{G.}~\bibnamefont{Dresselhaus}},
  \bibnamefont{and}
  \bibinfo{author}{\bibfnamefont{M.}~\bibnamefont{Dresselhaus}},
  \bibinfo{journal}{Phys. Rev. B} \textbf{\bibinfo{volume}{54}},
  \bibinfo{pages}{17954} (\bibinfo{year}{1996}).

\bibitem[{\citenamefont{Fujita et~al.}(1997)\citenamefont{Fujita, Igami, and
  Nakada}}]{Fujita97}
\bibinfo{author}{\bibfnamefont{M.}~\bibnamefont{Fujita}},
  \bibinfo{author}{\bibfnamefont{M.}~\bibnamefont{Igami}}, \bibnamefont{and}
  \bibinfo{author}{\bibfnamefont{K.}~\bibnamefont{Nakada}},
  \bibinfo{journal}{J. Phys. Soc. Jpn.} \textbf{\bibinfo{volume}{66}},
  \bibinfo{pages}{1864} (\bibinfo{year}{1997}).

\bibitem[{\citenamefont{Wakabayashi et~al.}(1998)\citenamefont{Wakabayashi,
  Sigrist, and Fujita}}]{Wakabayashi98}
\bibinfo{author}{\bibfnamefont{K.}~\bibnamefont{Wakabayashi}},
  \bibinfo{author}{\bibfnamefont{M.}~\bibnamefont{Sigrist}}, \bibnamefont{and}
  \bibinfo{author}{\bibfnamefont{M.}~\bibnamefont{Fujita}},
  \bibinfo{journal}{J. Phys. Soc. Jpn.} \textbf{\bibinfo{volume}{67}},
  \bibinfo{pages}{2089} (\bibinfo{year}{1998}).

\bibitem[{\citenamefont{Miyamoto et~al.}(1999)\citenamefont{Miyamoto, Nakada,
  and Fujita}}]{Miyamoto99}
\bibinfo{author}{\bibfnamefont{Y.}~\bibnamefont{Miyamoto}},
  \bibinfo{author}{\bibfnamefont{K.}~\bibnamefont{Nakada}}, \bibnamefont{and}
  \bibinfo{author}{\bibfnamefont{M.}~\bibnamefont{Fujita}},
  \bibinfo{journal}{Phys. Rev. B} \textbf{\bibinfo{volume}{59}},
  \bibinfo{pages}{9858} (\bibinfo{year}{1999}).

\bibitem[{\citenamefont{Harigaya}(2001{\natexlab{a}})}]{Harigaya01}
\bibinfo{author}{\bibfnamefont{K.}~\bibnamefont{Harigaya}},
  \bibinfo{journal}{J. Phys. Condens. Matter} \textbf{\bibinfo{volume}{13}},
  \bibinfo{pages}{1295} (\bibinfo{year}{2001}{\natexlab{a}}).

\bibitem[{\citenamefont{Harigaya}(2001{\natexlab{b}})}]{Harigaya01-2}
\bibinfo{author}{\bibfnamefont{K.}~\bibnamefont{Harigaya}},
  \bibinfo{journal}{Chem. Phys. Lett.} \textbf{\bibinfo{volume}{340}},
  \bibinfo{pages}{123} (\bibinfo{year}{2001}{\natexlab{b}}).

\bibitem[{\citenamefont{Okada and Oshiyama}(2001)}]{Okada01}
\bibinfo{author}{\bibfnamefont{S.}~\bibnamefont{Okada}} \bibnamefont{and}
  \bibinfo{author}{\bibfnamefont{A.}~\bibnamefont{Oshiyama}},
  \bibinfo{journal}{Phys. Rev. Lett.} \textbf{\bibinfo{volume}{87}},
  \bibinfo{pages}{146803} (\bibinfo{year}{2001}).

\bibitem[{\citenamefont{Harigaya and Enoki}(2002)}]{Harigaya02}
\bibinfo{author}{\bibfnamefont{K.}~\bibnamefont{Harigaya}} \bibnamefont{and}
  \bibinfo{author}{\bibfnamefont{T.}~\bibnamefont{Enoki}},
  \bibinfo{journal}{Chem. Phys. Lett.} \textbf{\bibinfo{volume}{351}},
  \bibinfo{pages}{128} (\bibinfo{year}{2002}).

\bibitem[{\citenamefont{Land et~al.}(1992)\citenamefont{Land, Micherly, Behm,
  Hemminger, and Comsa}}]{Land}
\bibinfo{author}{\bibfnamefont{T.}~\bibnamefont{Land}},
  \bibinfo{author}{\bibfnamefont{T.}~\bibnamefont{Micherly}},
  \bibinfo{author}{\bibfnamefont{R.}~\bibnamefont{Behm}},
  \bibinfo{author}{\bibfnamefont{J.}~\bibnamefont{Hemminger}},
  \bibnamefont{and} \bibinfo{author}{\bibfnamefont{G.}~\bibnamefont{Comsa}},
  \bibinfo{journal}{Surf. Sci.} \textbf{\bibinfo{volume}{264}},
  \bibinfo{pages}{261} (\bibinfo{year}{1992}).

\bibitem[{\citenamefont{Nagashima et~al.}(1994)\citenamefont{Nagashima, Itoh,
  Ichikawa, Oshima, and Otani}}]{Nagashima}
\bibinfo{author}{\bibfnamefont{A.}~\bibnamefont{Nagashima}},
  \bibinfo{author}{\bibfnamefont{H.}~\bibnamefont{Itoh}},
  \bibinfo{author}{\bibfnamefont{T.}~\bibnamefont{Ichikawa}},
  \bibinfo{author}{\bibfnamefont{C.}~\bibnamefont{Oshima}}, \bibnamefont{and}
  \bibinfo{author}{\bibfnamefont{S.}~\bibnamefont{Otani}},
  \bibinfo{journal}{Phys. Rev. B} \textbf{\bibinfo{volume}{50}},
  \bibinfo{pages}{4756} (\bibinfo{year}{1994}).

\bibitem[{\citenamefont{Thess et~al.}(1996)\citenamefont{Thess, LEe, Nikolaev,
  Dai, Petit, Robert, Xu, Lee, Kim, Rinzler et~al.}}]{Thess}
\bibinfo{author}{\bibfnamefont{A.}~\bibnamefont{Thess}},
  \bibinfo{author}{\bibfnamefont{R.}~\bibnamefont{LEe}},
  \bibinfo{author}{\bibfnamefont{P.}~\bibnamefont{Nikolaev}},
  \bibinfo{author}{\bibfnamefont{H.}~\bibnamefont{Dai}},
  \bibinfo{author}{\bibfnamefont{P.}~\bibnamefont{Petit}},
  \bibinfo{author}{\bibfnamefont{J.}~\bibnamefont{Robert}},
  \bibinfo{author}{\bibfnamefont{C.}~\bibnamefont{Xu}},
  \bibinfo{author}{\bibfnamefont{Y.}~\bibnamefont{Lee}},
  \bibinfo{author}{\bibfnamefont{S.}~\bibnamefont{Kim}},
  \bibinfo{author}{\bibfnamefont{A.}~\bibnamefont{Rinzler}},
  \bibnamefont{et~al.}, \bibinfo{journal}{Science}
  \textbf{\bibinfo{volume}{273}}, \bibinfo{pages}{483} (\bibinfo{year}{1996}).

\bibitem[{\citenamefont{Lee et~al.}(1997)\citenamefont{Lee, Kim, and
  Tom\'{a}nek}}]{Lee}
\bibinfo{author}{\bibfnamefont{Y.}~\bibnamefont{Lee}},
  \bibinfo{author}{\bibfnamefont{S.}~\bibnamefont{Kim}}, \bibnamefont{and}
  \bibinfo{author}{\bibfnamefont{D.}~\bibnamefont{Tom\'{a}nek}},
  \bibinfo{journal}{Phys. Rev. Lett.} \textbf{\bibinfo{volume}{78}},
  \bibinfo{pages}{2393} (\bibinfo{year}{1997}).

\bibitem[{\citenamefont{Klein}(1994)}]{Klein}
\bibinfo{author}{\bibfnamefont{D.~J.} \bibnamefont{Klein}},
  \bibinfo{journal}{Chem. Phys. Lett.} \textbf{\bibinfo{volume}{217}},
  \bibinfo{pages}{261} (\bibinfo{year}{1994}).

\bibitem[{\citenamefont{Klein and Bytautas}(1999)}]{Klein-Bytautas}
\bibinfo{author}{\bibfnamefont{D.~J.} \bibnamefont{Klein}} \bibnamefont{and}
  \bibinfo{author}{\bibfnamefont{L.}~\bibnamefont{Bytautas}},
  \bibinfo{journal}{J. Phys. Chem. A} \textbf{\bibinfo{volume}{103}},
  \bibinfo{pages}{5196} (\bibinfo{year}{1999}).

\bibitem[{\citenamefont{Lieb}(1989)}]{Lieb}
\bibinfo{author}{\bibfnamefont{E.}~\bibnamefont{Lieb}}, \bibinfo{journal}{Phys.
  Rev. Lett.} \textbf{\bibinfo{volume}{64}}, \bibinfo{pages}{1201}
  (\bibinfo{year}{1989}).

\bibitem[{\citenamefont{Wakabayashi and Sigrist}(2000)}]{Wakabayashi}
\bibinfo{author}{\bibfnamefont{K.}~\bibnamefont{Wakabayashi}} \bibnamefont{and}
  \bibinfo{author}{\bibfnamefont{M.}~\bibnamefont{Sigrist}},
  \bibinfo{journal}{Phys. Rev. Lett.} \textbf{\bibinfo{volume}{84}},
  \bibinfo{pages}{3390} (\bibinfo{year}{2000}).

\bibitem[{\citenamefont{Kusakabe and Takagi}(2002)}]{Kusakabe-Takagi}
\bibinfo{author}{\bibfnamefont{K.}~\bibnamefont{Kusakabe}} \bibnamefont{and}
  \bibinfo{author}{\bibfnamefont{Y.}~\bibnamefont{Takagi}},
  \bibinfo{journal}{to appear in Mol. Cryst. Liq. Cryst.}
  (\bibinfo{year}{2002}).

\bibitem[{\citenamefont{Kusakabe and Maruyama}(2002)}]{Kusakabe-Maruyama}
\bibinfo{author}{\bibfnamefont{K.}~\bibnamefont{Kusakabe}} \bibnamefont{and}
  \bibinfo{author}{\bibfnamefont{M.}~\bibnamefont{Maruyama}},
  \bibinfo{journal}{to appear in TANSO (in Japanese)}  (\bibinfo{year}{2002}).

\bibitem[{\citenamefont{Hohenberg and Kohn}(1964)}]{Hohenberg-Kohn}
\bibinfo{author}{\bibfnamefont{P.}~\bibnamefont{Hohenberg}} \bibnamefont{and}
  \bibinfo{author}{\bibfnamefont{W.}~\bibnamefont{Kohn}},
  \bibinfo{journal}{Phys. Rev.} \textbf{\bibinfo{volume}{136}},
  \bibinfo{pages}{B864} (\bibinfo{year}{1964}).

\bibitem[{\citenamefont{Kohn and Sham}(1965)}]{Kohn-Sham}
\bibinfo{author}{\bibfnamefont{W.}~\bibnamefont{Kohn}} \bibnamefont{and}
  \bibinfo{author}{\bibfnamefont{L.}~\bibnamefont{Sham}},
  \bibinfo{journal}{Phys. Rev.} \textbf{\bibinfo{volume}{140}},
  \bibinfo{pages}{A1133} (\bibinfo{year}{1965}).

\bibitem[{\citenamefont{Perdew and Wang}(1992)}]{LDAPW92}
\bibinfo{author}{\bibfnamefont{J.~P.} \bibnamefont{Perdew}} \bibnamefont{and}
  \bibinfo{author}{\bibfnamefont{Y.}~\bibnamefont{Wang}},
  \bibinfo{journal}{Phys. Rev. B} \textbf{\bibinfo{volume}{45}},
  \bibinfo{pages}{13244} (\bibinfo{year}{1992}).

\bibitem[{\citenamefont{Vanderbilt}(1990)}]{Vanderbilt}
\bibinfo{author}{\bibfnamefont{D.}~\bibnamefont{Vanderbilt}},
  \bibinfo{journal}{Phys. Rev. B} \textbf{\bibinfo{volume}{41}},
  \bibinfo{pages}{7892} (\bibinfo{year}{1990}).

\bibitem[{\citenamefont{Yamauchi et~al.}(1996)\citenamefont{Yamauchi, Tsukada,
  Watanabe, and Sugino}}]{Yamauchi}
\bibinfo{author}{\bibfnamefont{J.}~\bibnamefont{Yamauchi}},
  \bibinfo{author}{\bibfnamefont{M.}~\bibnamefont{Tsukada}},
  \bibinfo{author}{\bibfnamefont{S.}~\bibnamefont{Watanabe}}, \bibnamefont{and}
  \bibinfo{author}{\bibfnamefont{O.}~\bibnamefont{Sugino}},
  \bibinfo{journal}{Phys. Rev. B} \textbf{\bibinfo{volume}{54}},
  \bibinfo{pages}{5586} (\bibinfo{year}{1996}).

\bibitem[{\citenamefont{Furthmuller et~al.}(1994)\citenamefont{Furthmuller,
  Hafner, and Kresse}}]{Furthmuller}
\bibinfo{author}{\bibfnamefont{J.}~\bibnamefont{Furthmuller}},
  \bibinfo{author}{\bibfnamefont{J.}~\bibnamefont{Hafner}}, \bibnamefont{and}
  \bibinfo{author}{\bibfnamefont{G.}~\bibnamefont{Kresse}},
  \bibinfo{journal}{Phys. Rev. B} \textbf{\bibinfo{volume}{50}},
  \bibinfo{pages}{15606} (\bibinfo{year}{1994}).

\bibitem[{\citenamefont{Kertesz and Hoffmann}(1983)}]{Kertesz}
\bibinfo{author}{\bibfnamefont{M.}~\bibnamefont{Kertesz}} \bibnamefont{and}
  \bibinfo{author}{\bibfnamefont{R.}~\bibnamefont{Hoffmann}},
  \bibinfo{journal}{Solid State Commun.} \textbf{\bibinfo{volume}{47}},
  \bibinfo{pages}{97} (\bibinfo{year}{1983}).

\bibitem[{\citenamefont{Maruyama and Kusakabe}(2002)}]{Maruyama-Kusakabe}
\bibinfo{author}{\bibfnamefont{M.}~\bibnamefont{Maruyama}} \bibnamefont{and}
  \bibinfo{author}{\bibfnamefont{K.}~\bibnamefont{Kusakabe}},
  \bibinfo{journal}{unpublished}  (\bibinfo{year}{2002}).

\end{thebibliography}

\end{document}